\documentclass[3p,twocolumn]{elsarticle}

\usepackage[version=3]{mhchem} 
\usepackage[T1]{fontenc}     
\usepackage{verbatim}
\usepackage{float}
\usepackage{subfig}
\usepackage{graphicx}
\usepackage{color}
\usepackage{xcolor}
\usepackage{physics}
\usepackage{lineno,hyperref}
\usepackage{siunitx}
\modulolinenumbers[5]

\newcommand{\fig}[1]{Fig.~\ref{#1}}

\newcommand{\myeqref}[1]{Eq.~(\ref{#1})}
\newcommand{\myeqrefFromTo}[2]{Eq.~(\ref{#1})-(\ref{#2})}

\newcommand{\mysubsectionstar}[1]{\subsection*{#1}}
\newcommand{\myfigwidth}{.48\textwidth}
\newcommand{\artof} {ArTOF }
\newcommand{\artofs}{ArTOFs }
\newcommand{\degree}{$^\circ$}
\newcommand{\ogamma}{\tfrac{1}{\gamma}}
\newcommand{\mycomment}[1]{}

\newcommand{		\figCoescaSetup	}{
\begin{figure}
	\centering
	\includegraphics[width=\myfigwidth]{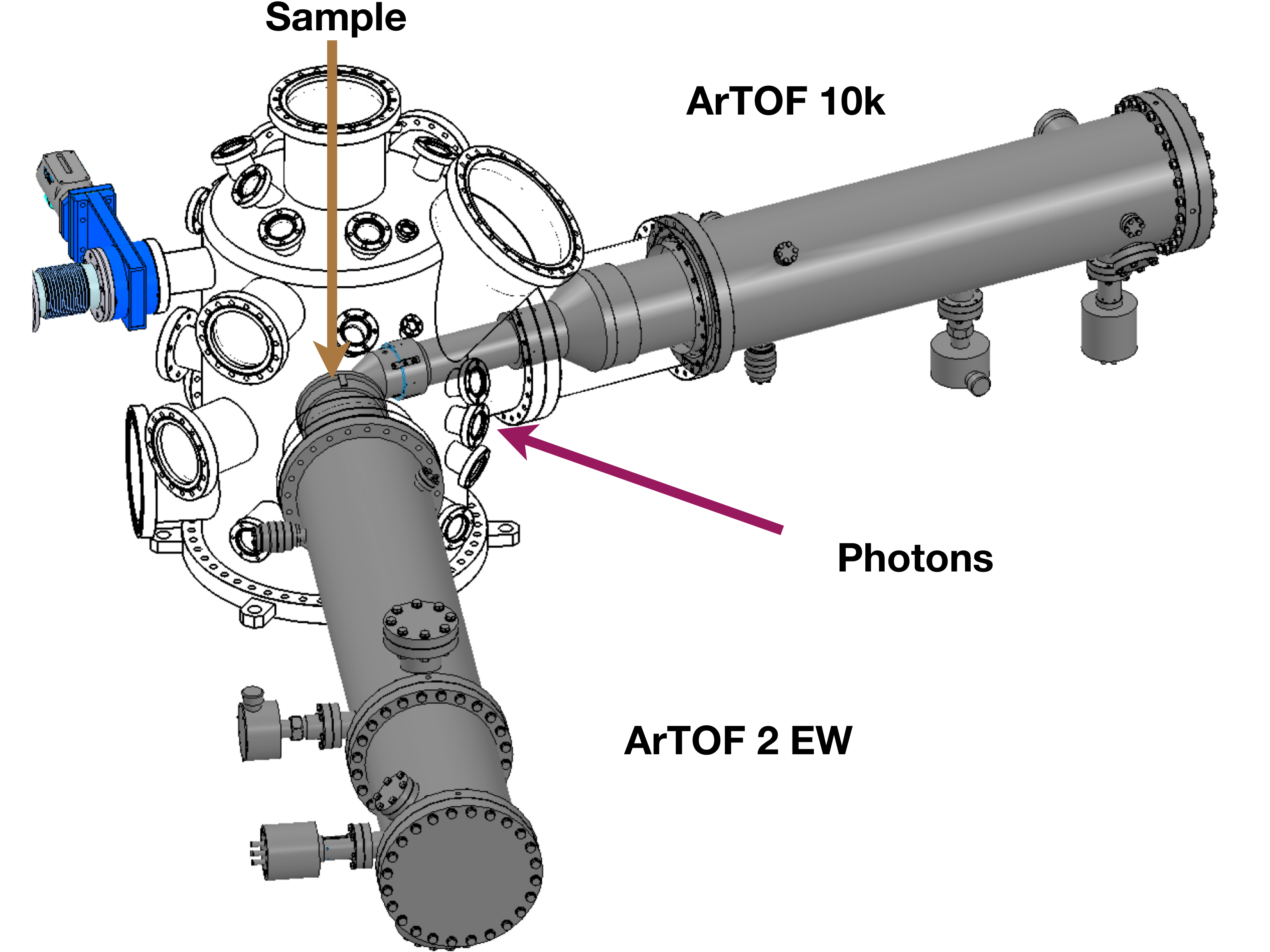}
	\caption{Schematic of the CoESCA measurement setup. The X-rays enter the experiment from the right side in the figure and hit the sample in front of the spectrometers. The sample is not shown in this illustration. It is mounted on a 5-axis manipulator installed at the top flange of the analysis chamber.}
	\label{fig:CoESCA}
\end{figure}
}

\newcommand{		\figCoAvsToA	}{
\begin{figure}
	\centering
	\includegraphics[width=\myfigwidth]{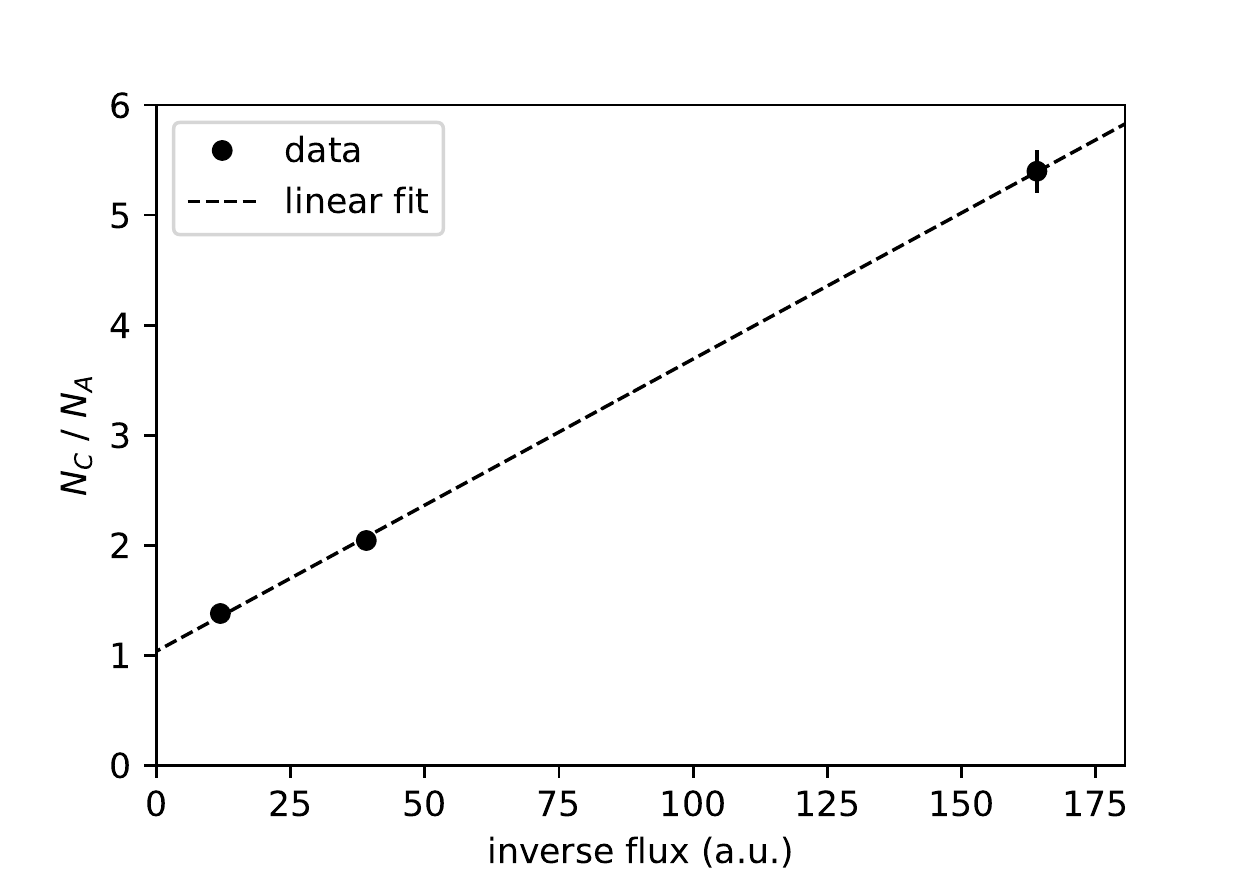}
	\caption{
Ratio R of coincidence counts from one pulse $N_{coin}$ and coincidence counts from two pulses $N_{acc}$  versus the inverse flux. The fitted line extrapolates to $R_0=1.038 \pm 0.001$ for an inverse flux of 0, which confirms the validity of our assumptions in \myeqrefFromTo{eq:Poisson}{eq:NcOverNa}.
	}
	\label{fig:CoA_vs_ToA}
\end{figure}}

\newcommand{		\figSNReff	}{
\begin{figure}
	\centering
	\includegraphics[width=\myfigwidth]{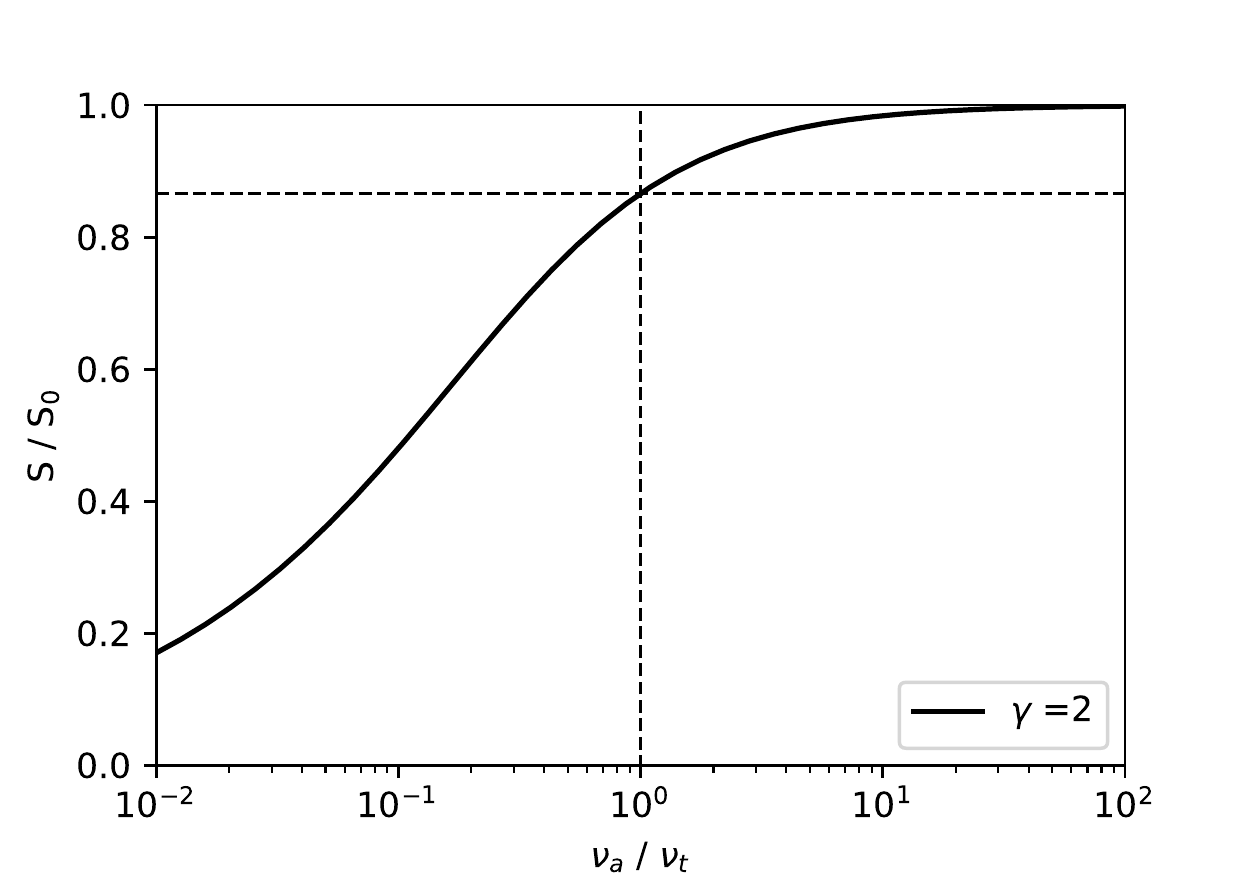}
	\caption{Effective signal-to-noise ratio $S / S_0$ versus photon flux in units of $\nu_a / \nu_t$. The dashed lines mark $\nu_a / \nu_t=1$. See text for more details.}
	\label{fig:nueff_vs_logflux}
\end{figure}}

\newcommand{		\figCoinAndAccMap	}{
\begin{figure}[t]
	\centering
 	\subfloat{ \includegraphics[width=0.45\textwidth] {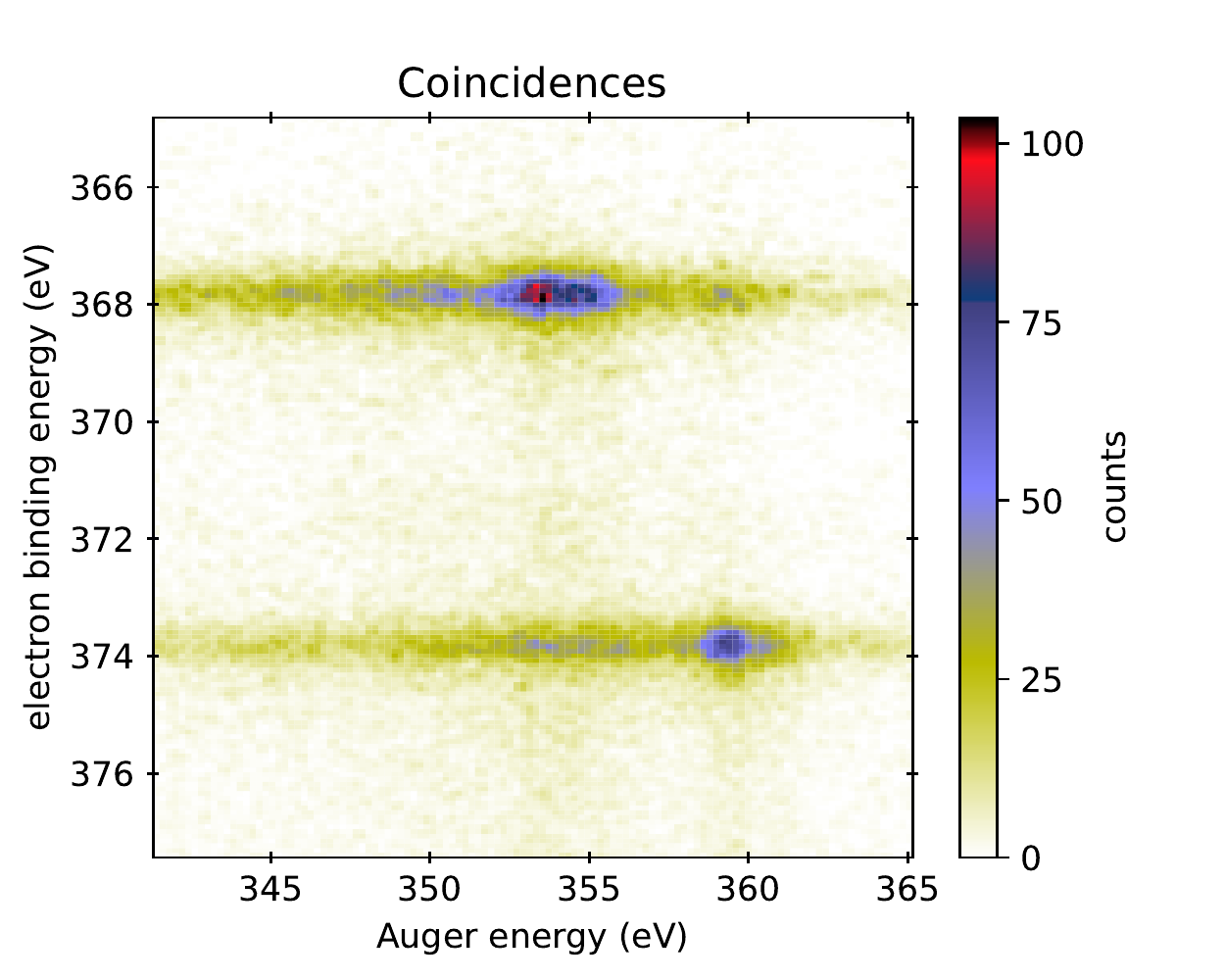} }
	\\ \vspace{-2em}
	\subfloat{ \includegraphics[width=0.45\textwidth] {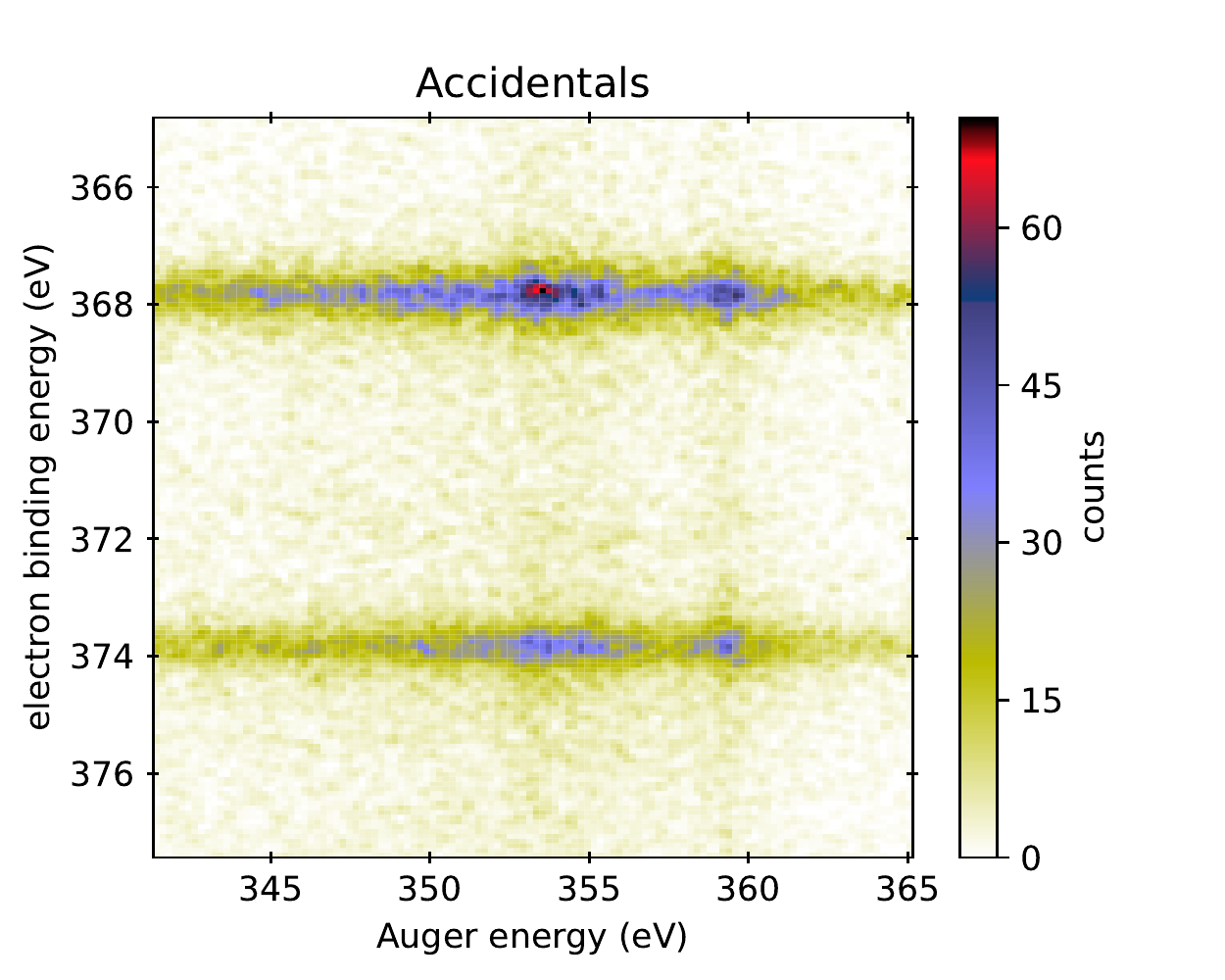} }
	\caption{Raw data maps (2D-histograms): \emph{(top)} All coincidences acquired from single pulses. \emph{(bottom)} Accidentals recorded for two consecutive pulses. The energy of the Auger electron is plotted on the x-axis and the binding energy of the photoelectron is shown on the y-axis.}
	\label{fig:rawmaps}
\end{figure}
}

\newcommand{		\figTrueMapWithSpecs	}{
\begin{figure}[t]
	\centering
	\includegraphics[width=0.45\textwidth] {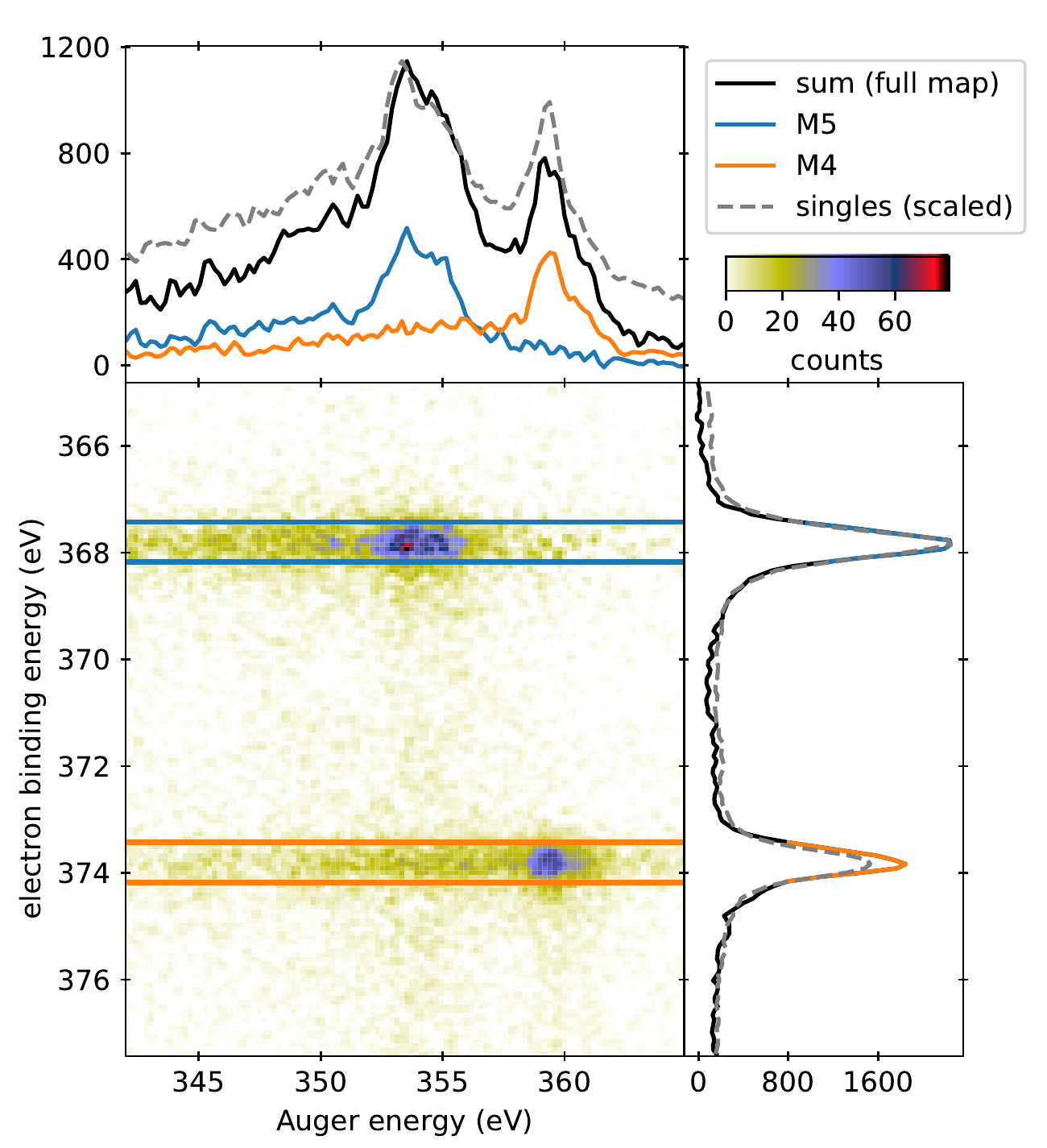}
	\caption{True coincidence map derived from the raw data shown in \fig{fig:rawmaps}, see text for more details on how the true map is obtained. 
		     Colored frames in the map indicate the integration regions for the Auger spectra in coincidence with the M$_4$ (3d$_{3/2}$, orange) or the M$_5$ (3d$_{5/2}$, blue), respectively.
		     On the left are the integrated photoelectron spectra and the Auger spectra are plotted on top of the map. 
		     In addition, the sum of the true map (black) and the non-coincident singles spectra (grey, dashed) derived from the accidentals map are shown. 
		     The M$_5$NN Auger spectrum is a intrinsic spectrum already, while the M$_4$NN spectrum still contains up to 6\% of true coincident background originating from scattered M$_5$ photoelectrons. 
		     See text and \fig{fig:specsAuger_cleanANDtwoholeaxis} for a description of the origin of this background and an estimated intrinsic M$_4$NN Auger spectrum.
		}
	\label{fig:truemap}
\end{figure}
}

\newcommand{		\figSpecAugerAndTwoHoleAxis		}{
 \begin{figure}[t]
	\centering
	\includegraphics[width=.45\textwidth]{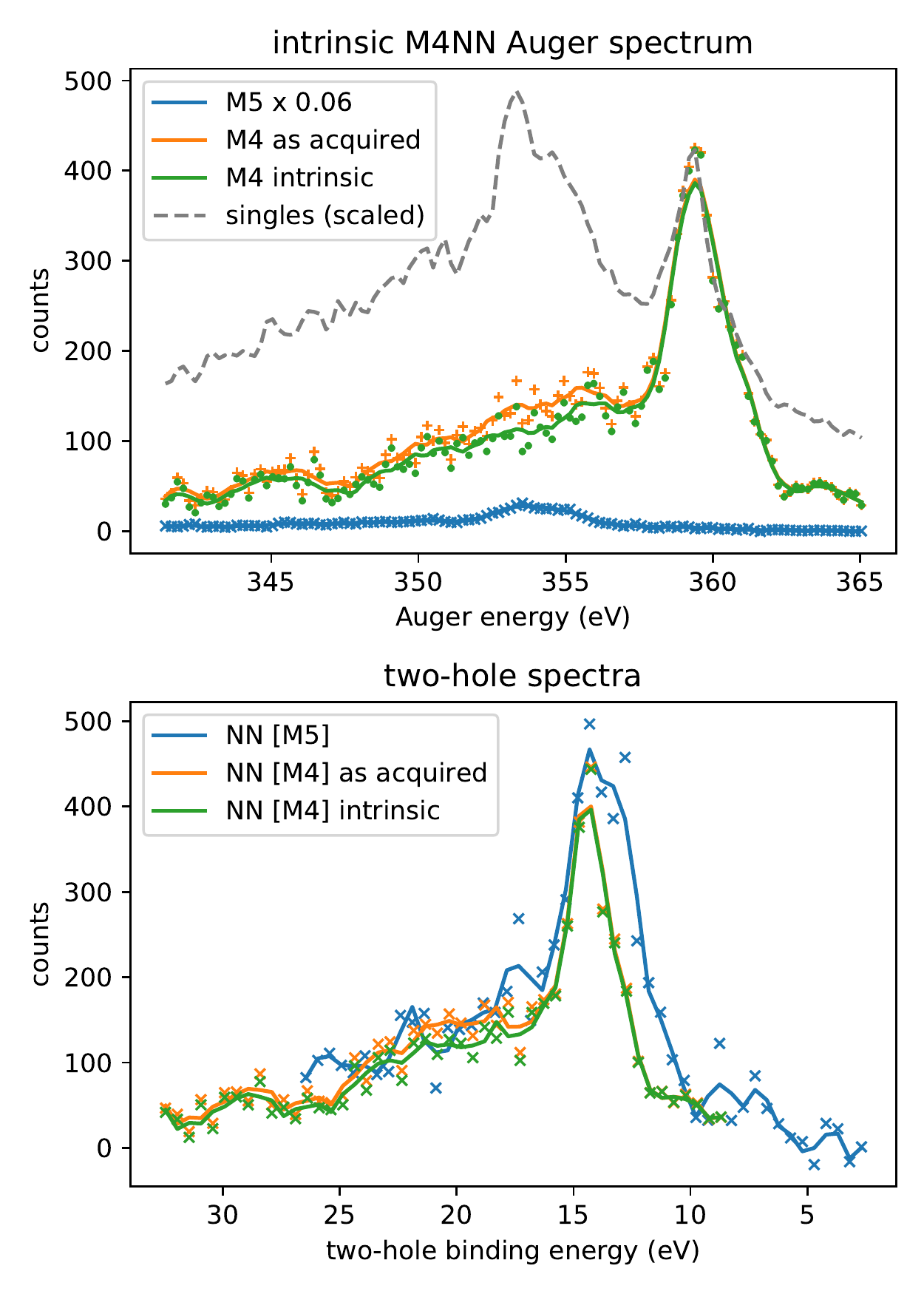}
	\caption{  \emph{(top) }Measured M$_4$VV coincidence spectrum (orange), scaled M$_5$NN coincidence spectrum (blue) and their difference as best guess for the intrinsic M$_4$NN spectrum (green, see text for details). 
	            \emph{ (bottom)} Two-hole binding energy spectra: Comparison of M$_5$NN (blue), M$_4$NN as measured (orange) and derived intrinsic M$_4$NN (green).
	            The markers in both plots are the measured data. 
	            The solid lines are the data smoothed with a Savitzky-Golay filter \cite{savgol} as guide for the eye.}
	\label{fig:specsAuger_cleanANDtwoholeaxis}
 \end{figure}
}
\journal{Journal of Electron Spectroscopy and Related Phenomena}

\begin{document}

\begin{frontmatter}

\title{The CoESCA station at BESSY: Auger electron --- Photoelectron coincidences from surfaces demonstrated for Ag MNN }

\author[ubjl,hzb]{T. Leitner} \corref{cor1}
\cortext[cor1]{Corresponding author} \ead{torsten.leitner@gmail.com}
\author[hzb,potsdam]{A. Born}
\author[ubjl,hzb]{I. Bidermane}
\author[ubjl,hzb]{R. Ovsyannikov}
\author[uu]{F. O. L. Johansson}
\author[chalmers]{Y. Sassa}
\author[ubjl,hzb]{A. F\"ohlisch}
\author[ubjl,uu]{A. Lindblad}
\author[mpihalle]{F. O. Schumann}
\author[ubjl,uu]{S. Svensson}
\author[ubjl,uu]{N. M\aa rtensson}

\address[ubjl]{UBjL, Uppsala-Berlin joint Laboratory}
\address[hzb]{Institute for Methods and Instrumentation in Synchrotron Radiation Research, Helmholtz-Zentrum Berlin f\"ur Materialien und Energie, Albert-Einstein-Strasse 15, 12489 Berlin, Germany}
\address[potsdam]{Institut f\"ur Physik und Astronomie, Universit\"at Potsdam, Karl-Liebknecht-Strasse 24-25, 14476 Potsdam, Germany}
\address[uu]{Division Molecular and Condensed Matter Physics, Department of Physics and Astronomy, Uppsala University, Box 516, SE-751 20 Uppsala, Sweden}
\address[mpihalle]{Max-Planck-Institut f\"ur Mikrostrukturphysik, Weinberg 2, 06120 Halle, Germany}
\address[chalmers]{Materials Physics, Chalmers University of Technology, SE-412 96 Gothenburg, Sweden}

\begin{keyword}
Auger \sep Photoelectron  \sep coincidence  \sep spectroscopy \sep AES  \sep PES \sep APECS  \sep surface science
\end{keyword}

\begin{abstract}
In this work, we present the CoESCA station for electron-electron coincidence spectroscopy from surfaces, built in a close collaboration between Uppsala University and Helmholtz-Zentrum Berlin at the BESSY II synchrotron facility in Berlin, Germany. 
We start with a detailed overview of previous work in the field of electron-electron coincidences, before we describe the CoESCA setup and its design parameters.
The system is capable of recording shot-to-shot resolved 6D coincidence datasets, i.e. the kinetic energy and the two take off angles for both coincident electrons.
The mathematics behind extracting and analysing these multi-dimensional coincidence datasets is introduced, with a focus on coincidence statistics, resulting in fundamental limits of the signal-to-noise ratio and its implications for acquisition times and the size of the raw data stream.
The functionality of the CoESCA station is demonstrated for the example of Auger electron --- photoelectron coincidences from silver surfaces for photoelectrons from the Ag 3d core levels and their corresponding  MNN Auger electrons.
The Auger spectra originating from the different core levels, $3d_{3/2}$ and $3d_{5/2}$ could be separated and further, the two-hole state energy distributions were determined for these Auger decay channels.
\end{abstract}

\end{frontmatter}

\section{Introduction}
In 1923 Pierre Auger discovered the effect named after him. However, already in 1922 Lise Meitner made a comment about such a phenomenon when explaining some details in a beta radiation spectrum. For Meitner this was just a small addition to the main processes she was studying. Auger, on the other hand, dedicated his whole doctoral thesis to exploring this phenomenon, see for example \cite{DUP78}. 
After core hole photoionisation, an Auger process can ocure, where the core hole is filled by an electron from a less bound shell, while a second electron is emitted, carrying the excess energy. 
A two-hole state is produced and two electrons linked to each other leave the sample, an Auger electron and a photoelectron.
Since the initial vacancy state and the final two-hole state have well-defined energies for each element in the periodic table, the Auger electrons will be emitted with characteristic energies. 
In this way Auger Electron Spectroscopy (AES) can be used to identify which elements are present in a sample. Furthermore, for Auger transitions involving shallow core-levels, the Auger energies will typically be in the range of a few hundred eV. For such electron energies the mean free path will be relatively short and the Auger spectral information will be surface sensitive. For this reason AES became a very powerful and much used tool in surface science.

One of the great successes of photoelectron spectroscopy is that one can perform detailed chemical analysis using the chemical shift. 
Also Auger transitions involving two core levels show chemical shifts. 
Due to the different, and often much larger chemical shifts, AES provides additional opportunities to distinguish different chemical species compared to photoelectron spectroscopy. 
Since the chemical shifts are different for Auger transitions and for core level photoelectrons, also the difference in kinetic energy between these features will undergo a chemical shift. 
This difference, denoted the Auger parameter,  \cite{WAG72,WAG88} has also been used for analytical purposes. The main reason for using the Auger parameter in an X-ray excited experiment is that it does not require any calibration of the kinetic energy scale. This is for instance an advantage for insulating samples where there may be sample charging.

An important aspect of AES is that it provides information about the two-hole states of a system. 
For this reason it is well suited for probing electron correlation effects.
In Core Valence Valence (CVV) Auger spectra from transition metals, it can be that the two holes are delocalised, then the two-hole spectrum is a self-convolution of the single hole valence spectrum, just as measured by photoelectron spectroscopy. 
However, when electron correlation effects become more important towards the end of a transition series, localised final states start appearing, where the two final state holes are localised at the same site. 
For late transition elements such as silver the spectrum is completely dominated by localised two-hole states.
These states can be separated and accurately probed by coincidence spectroscopy.

There are a number of factors which often make the detailed analysis of Auger spectra complicated.
All core levels are broadened due to their finite lifetime. The lifetime is typically in the range of a few fs and shorter. This leads to a broadening of the spectral features in the range of a few tenths of an eV up to a couple of eV for the transitions, which are most often used in Auger electron studies. In addition, also the two-hole final state has a finite lifetime. The lifetime broadening is more difficult to handle using fitting procedures for the Auger spectra than for photoelectron spectra. The two final state holes lead to multiplet effects and more complicated line shapes, which makes it more difficult to accurately parametrise the spectral profile. Often, there are also overlapping satellite transitions, due to multiple excited initial states caused by shake-up or shake-off processes in the creation of the initial vacancy. These multiple excited states may also be caused by the decay of deeper core hole states. These Auger satellites usually overlap the main spectral features, which severely complicates any detailed analysis.

One way to avoid these complications is to measure the Auger spectra in coincidence with the photoelectron spectra corresponding to the creation of the initial vacancy. This technique is usually denoted Auger PhotoElectron Coincidence Spectroscopy (APECS). In the present paper we describe the Coincidence Electron Spectroscopy for Chemical Analysis setup (CoESCA) at the BESSY II storage ring at HZB, Berlin. This system uses two high-resolution wide-angle time-of-flight \artof spectrometers. At first, we give a short overview of the field. Then we describe the CoESCA setup, and finally present some initial results.

\mysubsectionstar{Auger PhotoElectron Coincidence Spectroscopy - APECS}

There are a number of advantages of APECS, despite the fact that it involves quite demanding measurements. The key parameter is the ratio between true and accidental coincidences, which very much limits the useful count rates. The number of true coincidences will be proportional to the intensity of the exciting radiation, $I$. The number of accidental coincidences will however scale as $I^2$. This sets a limit on what intensities are practical to use. 
The way to increase this ratio is to use spectrometers with the highest possible electron acceptance angle. 
For this reason there are still only few examples of successful APECS measurements. 

In a pioneering paper, Haak et al.\cite{HAK78} used APECS to disentangle the different components of the Cu $L_{23} M_{45} M_{45}$ Auger spectrum. They identified satellite lines in the $L_3 M_{45} M_{45}$ part of the spectrum, caused by the $L_2 M_{45} M_{45}$ Coster-Kronig (CK) process, by performing measurements in coincidence with the 2p$_{1/2}$ core electron spectrum.

In gas phase measurements, different types of satellites have been identified \cite{ULR08}. 
For N$_2$, the participator Auger satellites were disentangled in this way \cite{NEE96}. For O$_2$, the contributions of the O$1s$ $^2\,\Sigma$ and $^4\,\Sigma$ exchange split states to the KVV spectrum could be resolved \cite{ARI12}. For CO it has furthermore been shown how one can study the Auger decay from vibrationally resolved core hole states \cite{ULR08}.

Thurgate and coworkers have reported a number of studies using APECS \cite{THU94, THU95, THU96,  LUN97}. In \cite{THU94} they compare the Ni APECS spectra for Ni and an NiFe alloy and identify changes in the Ni electronic states due to alloying. In Ref. \cite{LUN97} they investigate trends in the APECS spectra in 3d transition metals from Fe to Ga. They study the transition from bandlike final states in Fe and Co to localized final stets in Ga. They also use the technique to study the reduction of the $L_2L_3M_{45}$ Coster-Kronig rate going from Fe to Ga.

One important property of APECS is that it enables the distinction of Auger spectra from different chemically shifted components of a system. 
The Auger spectra are quite broad and the differently shifted components usually overlap. 
However, by measuring  the Auger spectrum in coincidence with chemically shifted, distinguishable, and often narrow, core-level peaks these spectra can be disentangled. 
In this way Auger spectra from different species of Al exposed to oxygen have been identified\cite{jensen1992coinmathsEtAl}. 
Furthermore, the surface core level shift for Ta(100) \cite{jensen1992coinmathsEtAl} and for the Ta4f level in TaC(111) \cite{BAR92} have been used to separate the Auger spectra originating from the bulk and the surface. 

There have also been several studies of Si surfaces and their oxidation. Auger spectra from different sites at the reconstructed Si(111)7x7 surface have been separated by APECS \cite{KAK07,KAK11} and the different stages of oxidation of this surface have been studied \cite{KAK12}. For an oxidized Si surface several oxidation stages occur at the same time. However, these can be well distinguished in XPS due to well-separated core level peaks. Using APECS it has thus been possible to record a pure Auger spectrum of SiO$_2$ \cite{STE04}. It is in fact rather difficult to record a spectrum of pure SiO$_2$ in other ways since it may be difficult to achieve homogenous samples. Bulk SiO$_2$ may decompose at the surface and for thicker samples there may also be charging effects, which will affect the spectra.

One useful property of APECS, when studying solids, is that the coincident spectra are more surface sensitive than the single photoelectron or Auger electron spectra. This is due to the fact that both the photoelectron and the Auger electron signals are attenuated due to the finite mean free path, which implies that the coincidence signal attenuates as a product of these effects. This increased surface sensitivity of APECS was for instance used to enhance the Si2p signal from the outermost Si atoms for the reconstructed Si(100) 2x1 surface \cite{KAK10}. 
Using APECS it is also possible to tune the surface sensitivity for the Auger spectra.  The Auger electrons are measured in coincidence with the photoelectrons of different kinetic energies, hence photoelectrons with different mean free paths from different depths in the sample. Since the Auger electrons are measured in coincidence, they stem from the same depth. This possibility may be particularly interesting for liquids and similar systems where there are no other ways to modify the surface sensitivity for the Auger spectra, like for example by measuring at different takeoff angles. 

The facet of the surface sensitivity was explored in another study on an Al surface. When the 2p core level spectrum of Al(111) was measured in coincidence with the Auger signal, a 40\% decrease of the intensity of the bulk plasmon loss signal was observed, reflecting the reduced information depth  \cite{jensen1992coinmathsEtAl}.
The surface plasmon loss signal, on the other hand, was only marginally reduced as expected. The mechanism for the increased surface sensitivity has also been investigated for a Si(100) surface \cite{WER05}. As for Al(111) the plasmon signal decreases when the photoelectron spectrum is measured in coincidence with the LVV Auger signal. However, the photoelectron spectrum was also measured in coincidence with the first plasmon loss signal of the LVV Auger spectrum. The Auger electrons, which have undergone one plasmon loss, originated from a larger depth and hence a significantly smaller reduction of the photoelectron plasmon loss signal could be observed. The increased surface sensitivity has also been seen through the reduction of the Auger electron diffraction effects for a Ge(100) surface when the measurements were performed in coincidence with the photoelectrons \cite{gotter2003angle}.

APECS has also been used to study the electronic structure of the Cu(001)-p(2x2)S adsorbate system \cite{filippo2015}. 
The overlapping $L_2VV$ and $L_3VV$ spectra could be separated by measuring in coincidence with the S 2p$_{1/2}$ and 2p$_{3/2}$ photoelectron lines, respectively, and their intrinsic line shapes could be determined. 
Overlayers of sulfur on Cu(001) and Ni(001) surfaces at different coverages and thereby for different local environments of the S atoms have also been studied using APECS \cite{filippo2016}. 
Changes in the sulphur derived density of states around the Fermi energy were identified when moving from a copper to a nickel substrate.

When measuring Auger spectra in coincidence with the photoelectron spectra the uncertainty due to the finite lifetime can be removed. 
In such a case, an energy loss $-\Delta E$ for the photoelectron will by the conservation of energy lead to energy gain $\Delta E$ for the corresponding Auger electron. 
The only remaining lifetime broadening in the APECS spectra will hence be caused by the finite lifetime of the final states. 
This implies that one can study Auger processes and Auger two-hole final states which are not broadened by the lifetime of the core hole state, under the assumption that the two-hole state decay times are much longer than decay times of the selected initial core hole state, so that the selected initial state can decay independently \cite{ohno99learnfromAPECS}.
One advantage of this is that one is not restricted to using the narrowest core levels in order to achieve high resolution.

This narrowing effect was demonstrated for Cu where the 3p$_{3/2}$ line was measured in coincidence with the $M_3 M_{45} M_{45}$ Auger spectrum \cite{JEN89}. The 3$_{3/2}$ line has a width of 2.2 eV, while the  $M_{45} M_{45} {}^1G$ final state has a width of only 1 eV. When the 3p$_{3/2}$ spectrum was measured in coincidence with a fixed and narrow energy window for the Auger spectrum they measured a photoelectron peak, which was narrower than the 2.2 eV expected from its lifetime. The intrinsic lifetime of the 3p$_{3/2}$ level was instead manifested as a shift within this 2.2 eV span as the energy of the Auger window was moved. The line narrowing effects were also clearly demonstrated in an investigation of Xe \cite{VIE98}.

In this connection the $N_{23}N_{45}N_{45}$ spectra in the 4d transition elements and the following elements up to Xe constitute a particularly interesting case. 
The Super Coster-Kronig character of the $N_{23}N_{45}N_{45}$ decay makes it very rapid and a one-step description of the excitation-deexcitation process becomes absolutely crucial. 
The distinction between the photoelectron and the Auger electron becomes less relevant and the shapes of the individual spectra may be heavily distorted \cite{SvanteNils1976}.  
However, by determining the sum of the energies of the emitted electrons with APECS, one can still accurately probe the energy spectrum of the $N_{45}N_{45}$ double-hole states, as demonstrated for Pd \cite{Kostanovskiy2015} and Ag \cite{Wei2014}. 
For Pd, the 4p photoelectron peaks are broad with a Lorentzian width of 4.7 eV, but except for the additional Lorentzian broadening their line shapes are the same as for the narrower 3d lines \cite{Mar81_Ag_M4toM5ratio}. 
In these works it was demonstrated that a correlation map can reveal the nature of the Auger decay. In the case of the 4p decay of Pd and Ag surfaces it was found that the Auger decay is described best by a 1-step process, while the 3d decay proceeded in a two-step manner. Such a behaviour can also be expected for the elements following Ag in the periodic table on the basis of their heavily distorted  4p line shape.
The concept of first creating a 4p hole, which then decays, completely breaks down. 
In fact, it is no longer relevant to talk about $4p_{1/2}$ and $4p_{3/2}$ single hole states.
However, using the sum spectra, the  $N_{45}N_{45}$  double hole states can also in this case be accurately probed, and the spectra show a similar shape as when determined using the $M_{45}N_{45}N_{45}$ Auger decay \cite{Wei2014}.

At ELETTRA in Trieste, the coincidence setup (ALOISA) incorporates several spectrometers, which allows angle resolved coincidence measurements \cite{GOT01}. The angular capabilities have allowed the establishment of the technique denoted Dichroic Effects in Angle Resolved APECS (DEAR-APECS). This has for instance been used to separate the different multiplet terms in the $M_3 M_{45} M_{45}$ Auger spectrum from Cu(111) \cite{GOT09}. This technique has also been used to study antiferromagnetism in the NiO/Ag(001) \cite{GOT13}. Furthermore, in thin films of Fe on Cu(001) the dichroic effects have been used to identify spectral features where the holes are created either in the minority or in the majority bands \cite{GOT12}. It was then found that there were particularly strong correlation effects when the holes were created in the majority band.

One has also used coincidence techniques to probe the angular and energy correlation of a pair of electrons emitted from a surface upon impact of an electron of specific momentum \cite{SHU05} or excited by a beam of monochromatic photons \cite{SHU06}: In this way one has derived information on the exchange-correlation hole in solids. The two-hole emission spectrum has also been studied in direct double photoionization processes and at energies where Cu 3p holes can be created, showing similar energy sharing effects in both cases \cite{vRI10}.

\section{Experimental Setup}

\figCoescaSetup

The Coincidence Electron Spectroscopy for Chemical Analysis (CoESCA) end-station is built at the BESSY UE52-PGM beamline which is a conventional soft X-ray plane-grating monochromator beamline, installed at an elliptical APPLE-II-type undulator, with a combination of a cylindrical and spherical mirror to provide a small horizontal focus of about 100 $\mu$m near the monochromator exit slit. 
The available photon energy is in the range from 120 eV up to 1600 eV and  linear or elliptical polarisation can be set.
The beamline offers a photon flux in the order of 10$^{12}$ ph/s/100mA and a resolving power of more than 10000 at 400 eV photon energy \cite{coesca_webpage,Bat07}.

The CoESCA end-station is designed for  electron-electron coincidence measurements on either in-situ or ex-situ prepared samples.   
It consists of an analysis chamber, equipped with two angle resolved time of flight (ArTOF) spectrometers  \cite{ovsyannikov2013principles}, \fig{fig:CoESCA}. 
The two \artofs are of different types reflecting the evolution of this type of spectrometers. 
One is an \artof 10k (actually the prototype) with an acceptance angle of $\pm$14\degree and the other instrument is an \artof 2 EW type with a wide acceptance angle of $\pm$24\degree.
Both spectrometers are positioned at a 54.6\degree angle with respect to the incoming photon beam. 

The sample is placed on a 5-axis manipulator (xyz-position, polar- and azimuthal rotations) with active sample heating or cooling in a temperature range from 30 K up to 900 K. 
A radial sample transfer system connects the analysis chamber with a load lock for sample introduction, a sample storage and a sample preparation chamber. 
The preparation chamber is equipped with a 4-axis manipulator with resistive heating, direct current heating or electron bombardment options (up to 1200 K) and provides sample current and temperature, a sputter gun and a quartz balance. There are a number of free ports for further equipment like evaporators or gas dosing systems. 
Another utility chamber (not shown in \fig{fig:CoESCA}) is situated on top of the analysis chamber, and is equipped with mass spectrometer, sputter gun and extra flanges for more user equipment. 
In addition, there are free access ports to our sample transfer system for accommodating user provided chambers to carry out specific sample preparations. 

The \artof electron spectrometers use advanced electron lenses in combination with measuring the time-of-flight of the electron and its impact position on the detector at the end of the flight tube. 
From this, the exact particle trajectory through the lens system can be calculated including the take-off angle of the electron providing for full 3d electron distributions in "one shot", i.e. for one set of parameters for electron analyser, beamline and sample position. 
The kinetic energy is determined from the time-of-flight and the precisely known length of the flight path \cite{ovsyannikov2013principles}. 
The instruments can, of course, be operated in non-coincidence mode for advanced studies of materials where, for example, two different core level regions can be monitored simultaneously while scanning an experimental parameter, such as the sample temperature.

\artof spectrometers demand a pulsed light source with a minimum time separation between the pulses, which depends on the chosen electrostatic lens mode and the kinetic energy of the electrons under investigation. 
For \artof measurements at the UE52-PGM beamline, the pulse-picking-by-resonant-excitation (PPRE) mode of BESSY II is used, providing quasi single bunch pulses separated by 800 ns (repetition rate of 1.25 MHz) \cite{Hol14}. 
However, recent developments on the BESSY machine side give rise to  higher pulse repetition rates in the future.
They are developing a TRIBs mode of operation for the storage ring \cite{tribs1,tribs2}, where a selected number of the electron bunches is pushed to an alternative orbit.
The light pulses originating from the different orbits can be separated in the beamlines.
The experimentalist can set the beamline to the light from either orbit, enabling to effectively offer two pulse fill patterns at once.
One fill pattern can be optimised for one or a few experiments without effecting the larger majority of running experiments.
For the CoESCA station, at least up to 5 MHz pulses repetition rate, i.e. 4 times the present value, can be processed, yielding 4 times shorter acquisition times for the same data quality.

\section{Measuring Coincidences}
In our setup an event is stored as a coincidence, if one electron is detected in each spectrometer after excitation by the same photon pulse. 
The data is 7-dimensional:  3D for each electron (x,y-hit coordinates and arrival time coordinate t) stored together with a trigger timestamp:
\begin{equation}\label{eq:CoinEvent}
\text{Event} =  \{x_1,y_1,t_1,x_2,y_2,t_2,\text{\emph{trig}}\}.
\end{equation}

True coincidences are pairs of electrons which are created by absorption of one photon. 
However, there is also a background of accidental coincidences. 
In this case, the two detected electrons have been excited by the same photon pulse, but they originate from two different photons, hence different sites in the sample. 
The aim is to measure the true coincidence distribution as accurately as possible and in the shortest possible time. 
There is no way experimentally to directly distinguish true from accidental coincidences. 
However, one can instead separately determine the distribution of only accidental coincidences and subtract them to derive the distribution of true coincidences from the data. 
It is then essential that the distribution of accidental coincidences is measured in exactly the same way as the full coincidence distribution.

Our setup is ideally suited for achieving this. 
During the experiment, a database is written containing each event in each detector together with a timestamp, which serves as a unique photon pulse ID. 
In post-processing,  true \& accidental coincidences measured from the same pulse (all coincidences) are extracted from the data, for events where an electron was detected in both detectors for the same pulse.
Accidental coincidences are extracted for cases, where one electron is recorded in each spectrometer, 
but for events from two different photon pulses (accidental coincidences). 

The data lists are sorted into multi-dimensional histogram distribution matrices $M$. 
The distribution for coincidences from one photon and one site in the sample, the true coincidences dataset $M_t$, is derived in post-processing by a simple subtraction of the accidental distribution $M_a$ from the all coincidences distribution $M_c$:

\begin{equation}\label{eq:CalcTrueMap}
M_t = M_c - M_a 
\end{equation}
with
\begin{equation}\label{eq:CalcAccMap}
M_a = \ogamma \, M_\gamma,
\end{equation}

where $\gamma$ is the number of virtual pulses per real pulse, which are analysed for accidental coincidences to get the measured distribution of accidentals M$_\gamma$.
In our case, the accidental data is created from comparing  \textit{detector A, pulse N} with \textit{detector B, pulse N-1} and, vice versa, from comparing \textit{detector A, pulse N-1} with \textit{detector B, pulse N}. 
Hence, it is created from twice as many virtual pulses than the all coincidences map and therefore $\gamma=2$.

The full 6D coordinates can be used to analyse the data. In most cases one looks at 2D maps of $E_1$ vs. $E_2$ (see \fig{fig:rawmaps} and \fig{fig:truemap}), but any other combination of coordinates can be used, for example one can additionally discriminate by the angles of the electrons and look at coincidences resolved in energy- and k-space.

All events are stored in a list-like database together with a timestamp.
This enables a variety of possibilities for post treatment and analysis of the data.
One may for instance detect, that there have been some changes of the sample during the measurements due to contamination or radiation damage and that therefore only a subset of the collected data is usable. 
This can be based on a careful analysis of the singles spectra, extracted from the data base. 
The singles spectra are calculated by evaluating the signals for each \artof independently, without looking for coincidences.
This yields the same spectra as in a classic non-coincident setup. 
These spectra are very useful, since they usually have two to three orders of magnitude higher count rates than the coincidence data. 
If a subset is identified in the singles data, where the changes can be neglected, one knows that the corresponding subset can be used also for the coincidence data. 
Furthermore, when using only part of the coincidence data this can always be connected to the corresponding determination of the accidental coincidences. 
All these spectra have in this case been recorded under exactly identical conditions.

The fact that the events are stored together with a timestamp, makes the experiment well adapted to studies of time-dependent phenomena. 
One can study slow processes, where the theoretical time resolution is set by the separation of the x-ray pulses, which in the case of BESSY II in the single bunch mode is 800 ns. In praxis, this time resolution is much slower and governed by the data statistics, since the low count rates make it necessary to integrate over many x-ray pulses before enough useful information can be found in the data set.
For studies of more rapid processes one could use a femto- to picosecond pump laser pulse, with a tuneable delay to the x-ray pulse. 
In this case the time resolution is set by the length of the x-ray pulse, typically 40 ps at present. 
However, for now the CoESCA station is not equipped with such a pump laser system.

An Auger spectrum provides information on the energies of two-hole states of a system. 
Usually, the two-hole energies are determined from an Auger spectrum by subtracting the core level binding energy from the Auger energies.  By doing so, the Auger spectrum is plotted on an energy scale which corresponds to the binding energy for the two-hole final states. 
If, for a metallic solid, the core level binding energy and the Auger energies are referenced to the Fermi level, also the two-hole energy scale will be referenced to the Fermi level. 
This is a widely used method to obtain the two-hole spectrum, which is valid in regions, where there are no overlapping Auger transitions due to the decay of other core levels.
However, one has to accept, that the spectrum will be influenced by any broadening effect connected to the core ionisation. 
In particular, the life time broadening of the core level will be seen as a corresponding broadening of the Auger spectrum and hence of the corresponding two-hole spectrum.

However, in a coincidence measurement it is possible to derive directly the binding energy of the two-hole final state. 
Since, for each coincident event, the energies of the photoelectron and the Auger electron are both measured, also the sum of these energies can be determined. 
The two-hole energy is then obtained by subtracting this energy sum from the photon energy. 
In our set-up it is straight-forward to extract the two-hole energy spectrum. 
The two-hole spectrum can be derived directly from the true coincidence map by summing the kinetic energies for each point in the map. Alternatively, one can treat the data in the same way as when deriving the true coincidence map, only that one for each event is calculating the sum of the kinetic energies, for the coincidence recording as well as for the recording of the accidental coincidences.

The two-hole spectrum, derived in this way, is not broadened by the core-hole life time. 
In coincidence spectra recorded at high resolution this allows improved resolution for the two-hole spectra. 
The count rate, however, is lower for the coincidence data so in each case one has to judge which type of evaluation of the data provides the most accurate results.  
Furthermore, the direct derivation of the two-hole spectrum makes it possible to use also rather broad core levels in Auger spectroscopy. 
This is a great advantage since it extends the range of core levels which can be used for detailed measurements.


\mysubsectionstar{Effective coincidence count rate and limitations}

\mycomment{wording, formulas adapted}

The true and accidental coincidence count rates are small compared to the primary particle pulse repetition rate. 
For example, at CoESCA we typically see an order of $10^{-6}$ coincidence counts per photon pulse.
The measured signals from different pulses are pair-wise independent and do not influence each other.
Therefore, we can describe the behaviour of the probability of $k$ events detected for the same pulse for an average rate of events per pulse 
$\lambda$ with a Poisson distribution:
\begin{equation}\label{eq:Poisson}
P_\lambda(k) =  \frac{\lambda^k}{k!}\text{e}^{-\lambda}   \overset{(\lambda \ll 1)}{\;\approx\;}  \ \frac{\lambda^k}{k!} 
\end{equation}

$\lambda$ scales linearly with the incident flux, i.e. the number of primary particles per pulse (e.g. photons).
In \cite{jensen1992coinmathsEtAl}, Jensen et al. provide a detailed analysis of the link between electron count rates and experimental parameters, such as primary particle flux, total detection volume, detection efficiencies and cross-sections for electron pair creation.

For a true coincidence, one primary particle is absorbed ($k=1$), which creates one electron pair, leading to one coincidence event in the detectors. An accidental coincidence occurs, when two primary particles are absorbed in the sample ($k=2$), and two electrons, which are recorded by the detectors, are created at two different sites.
Events for more than two absorbed primary particles can be neglected, since the probability drops very fast with $(\lambda^k/k!)$.
The probabilities for detecting a true  ($k=1$) or accidental ($k=2$) coincidence   read as:
\begin{equation}\label{eq:TandAscaling}
 \begin{aligned}
P_\lambda(1) &\approx   \lambda , \\
P_\lambda(2) &\approx \frac{1}{2} \lambda^2. \\[1ex]
 \end{aligned}
\end{equation}

The probability of finding any number of events for a single pulse is the sum over $P_\lambda(k)$ for all $k$.
And the probabilities for any two distinct pulses $A$ \& $B$ are linearly independent.
Thus, the probability to record an accidental coincidence from these two pulses is the product of the probabilities of measuring one or more events from each pulse independently:
\begin{equation}\label{eq:Afrom2pulses}
 \begin{aligned}
 P^{A\&B}_\lambda &= P^{A}_\lambda \; P^{B}_\lambda= \left[ \sum_{k>0} P_\lambda(k)  \right]^2 \\
 &\approx \left[ P_\lambda(1)\right]^2 = \lambda^2 = 2 \; P_\lambda(2) \\
 \end{aligned}
\end{equation}

The equation above states, that the accidental coincidence signal, which is derived from comparing pairs of different pulses is twice as intense as the accidental contribution to the coincidences signal derived from single pulses only.
This is accounted for in \myeqref{eq:CalcAccMap} by the scaling factor $\ogamma$.

The true and accidental count rates are defined as counts per time. They are linearly proportional to the respective $P_\lambda(k)$. 
However, the proportionality constants are different for measuring true or accidental coincidences, 
  since the detection volume for accidental coincidences is larger than for true coincidences,
  and since there is a contribution to the accidental coincidences arising from competing events against which one is trying to discriminate, like for example inelastic scattering of higher energy electrons \cite{jensen1992coinmathsEtAl}.
In order to respect these proportionalities, constants are introduced, which connect the count rates $\nu_t$, $\nu_a$ with the respective probabilites $P_\lambda(k)$:

\begin{equation}\label{eq:PtoNu}
 \begin{aligned}
\nu_t  &= \sigma_t  \,\, P_\lambda(1) \propto   \lambda , \\
\nu_a &= \sigma_a \, P_\lambda(2) \propto    \lambda^2  . \\[1ex] 
 \end{aligned}
\end{equation}

The rate of all coincidence counts per time, measured from single pulses is given by the sum of the true and accidental coincidence count rates:
\begin{equation}\label{eq:TplusA}
\nu_{c} = \nu_{t} + \nu_{a}.
\end{equation}

For a given acquisition time $\tau$, the number of all coincidence counts measured in an experiment is $N_c = \nu_c \, \tau$ and the number of accidental counts in this signal is $N_a= \nu_a \, \tau$. 
The ratio of these counts can be expressed as:

\begin{equation}\label{eq:NcOverNa}
R = \frac{N_c}{N_a} = \frac{\nu_t + \nu_a}{ \nu_a}  \propto    \lambda^{-1} + 1
\end{equation}

and it approaches $R=1$ for an infinite flux of primary particles $\lambda$.

In order to validate the assumptions and approximations in \myeqrefFromTo{eq:Poisson}{eq:NcOverNa} for our setup, we performed short coincidence runs for three different light intensities and determined $N_c$ and $N_a$ for each run.
\fig{fig:CoA_vs_ToA} shows the ratios $R$ from these runs versus the inverse flux, which was deduced from the singles count rates.
A linear fit extrapolates $R_0=1.038 \pm 0.001$ for $\lambda^{-1}=0$, which is in good agreement with the theoretical value of 1 and validates our assumptions.
\figCoAvsToA

The ratio of accidental to true counts for events recorded from single pulses is an important parameter for coincidence experiments. From \myeqref{eq:PtoNu} we see that this ratio is given by:
\begin{equation}\label{eq:AoT2}
\nu_a / \nu_t \propto \lambda \propto \text{flux}. 
\end{equation}

The higher the flux, the higher the relative accidental background, therefore simply increasing the flux is not beneficial for a coincidence experiment. 
On the other hand, simply lowering the intensity in order to remove almost all accidental coincidences leads to low count rates and therefore long acquisition times.
In order to judge the flux and the acquisition time needed to achieve a dataset of acceptable quality,
 it is useful to look at the behaviour of the ratio of the derived true coincidences signal and its statistical error, i.e. the signal-to-noise ratio $S$ of the true coincidences dataset as calculated from the measured data: 

\begin{equation}\label{eq:S}
S = M_t / \Delta M_t
\end{equation}

This ratio can be interpreted as the amount of useful information, which is added to the data for a given acquisition time.
In \cite{jensen1992coinmathsEtAl}, the authors discuss the conditions for reaching a certain value of this ratio in a coincidence measurement, starting from very basic assumptions. 
In their considerations, the contributions to the statistical noise, which originate from the subtraction for calculating the true coincidences in \myeqref{eq:CalcTrueMap} are approximated as negligible. 
They find, that the signal-to-noise ratio approaches an asymptote, when the flux of primary particles becomes too high.
Here, we present a calculation of $S$, which is adapted to our data acquisition system and includes 
 all contributions to the statistical error of the derived true coincidences dataset, which arise from the measurements of all coincidences from single pulses, $M_c$, and the determination of the accidental coincidences $M_a$ from comparing pairs of different pulses.
These two datasets and their statistical noise can be defined in terms of the count rates $\nu_t$ and $\nu_a$ and the acquisition time $\tau$.
Please note, that the following considerations, about the statistical quality of the counts in our multidimensional data matrices $M_x$ apply to each single point within the matrices. 
Therefore, they also apply to the sum of a subset of these matrices, when binning the data.
The total count rates $N_x = \sum{M_x}$ are the extreme case of binning to one single point per dataset.

The coincidences originating from single pulses, $M_c$, are measured directly and can be described as:

\begin{equation}\label{eq:DeltaCoin}
 \begin{aligned}
           M_c &= (\nu_t + \nu_a) \tau  \\ 
 \Delta M_c &= \sqrt{M_{c} }  =  \sqrt{ (\nu_t + \nu_a) \tau}   \\
 \end{aligned}
\end{equation}

The accidental dataset $M_a$ is derived from $M_\gamma$, which was measured with a statistical error of $\Delta M_\gamma = \sqrt{M_\gamma}$, we get:

\begin{equation}\label{eq:DeltaAcc}
 \begin{aligned}
           M_a &= \ogamma M_\gamma  =  \nu_a \tau \\
 \Delta M_a &=
 \tfrac{\partial M_a}{\partial M_\gamma} \; \Delta M_\gamma = 
  \ogamma  \sqrt{M_\gamma}   =  \sqrt{ { \nu_a \tau} / {\gamma}}  \\
 \end{aligned}
\end{equation}

The true dataset $M_t$ and its statistical error can be written as:

\begin{equation}\label{eq:DeltaTrue}
 \begin{aligned}
 M_t &= M_c - M_a = \nu_t \tau  \\
 \Delta M_{t} &= \sqrt{ (\Delta M_{c} )^2 + (\Delta M_a)^2 } \\
   &=  \sqrt{ \left( \nu_t + (1+\ogamma) \nu_a \right) \tau }. \\ 
  \end{aligned}
\end{equation}

Hence, we get:

\begin{equation}\label{eq:SNR}
 \begin{aligned}
S   
 &=  \sqrt{ \frac{ \tau} {  \nu_t^{-1} + (1+\ogamma) \nu_a  \nu_t^{-2} } }    \\
 &\propto \sqrt{ \frac{\tau}{ \lambda^{-1} + 1 + {\tfrac{1}{\gamma}} }  }  \\  
 \end{aligned}
\end{equation}

with $\nu_t \propto \lambda$ and $\nu_a \propto \lambda^2$, see \myeqref{eq:PtoNu}. 

For infinite flux, $\lambda=\infty$, $S$ approaches an upper limit:

\begin{equation}\label{eq:SNR0}
 \begin{aligned}
S_0 = \lim\limits_{\lambda \to \infty}  S &=  
                 \sqrt{ \frac{ \tau} { ( 1 + {\tfrac{1}{\gamma}}  )  \nu_a  \nu_t^{-2} } }    \\
   &\propto  \sqrt{ \frac{\tau}{  1 + {\tfrac{1}{\gamma}}  } }   \\   
 \end{aligned}
\end{equation}

\figSNReff
\fig{fig:nueff_vs_logflux} plots the effective signal-to-noise ratio $S / S_0$ versus the photon flux in units of $\nu_a / \nu_t$ for  $\gamma=2$, which means that twice as many virtual pulses are used for the determination of the accidental dataset than real pulses have hit the sample.
Note, that the values for $S/S_0$ are only qualitative, since we neglected the proportionality constants for $\nu_t$ and $\nu_a$ in this representation.
However, we see that $S/S_0$ approaches its asymptote very fast and the gain in signal quality decreases rapidly already for increasing the flux above a level corresponding to an accidental to true ratio of $\nu_a/\nu_t=1$.
 Hence, the acquisition time, which is necessary to achieve a desired statistical quality of the true coincidence data does not substantially decrease anymore for going to higher fluxes.
On the other hand, a further increase of the flux quadratically increases the number of recorded events and hence the amount of data to be stored.
For example, at $\nu_A/\nu_T=10$, depending on the sample, the CoESCA station can easily produce more than 500 GB of coincidence data for a typical measurement week of $6 \times 12$-hour shifts.
This corresponds to two or three coincidence maps, i.e. combinations of Auger and Photoelectron energy window settings of the detectors.

In order to increase $S/S_0$, a data analysis scheme for working with $\gamma>2$ is currently being developed for the CoESCA station.
The accidentals dataset $M_\gamma$ will be created by adding up the accidentals analysis results from pulse pairs (N,N-1), (N,N-2), ..., (N,N-K), with $K = \gamma/2$. 
The theoretical limit of $S$ for ${\gamma \to \infty}$ corresponds to the intrinsic signal-to-noise ratio arising from the measurement of the coincidence signal only and equals to $M_t / \Delta M_c$.

The best way to strongly reduce acquisition times, while remaining at an acceptable data rate is to increase the number of experiments per time, hence the pulse repetition rate.
In this experiment, the pulse repetition rate from the storage ring was fixed to 1.25 MHz.
The earlier mentioned development of the TRIBs mode operation of the storage ring  \cite{tribs1,tribs2} could enable up to 5 MHz pulse repetition rate in the near future.

\section{Results \& Discussion}
Arena et al. have previously reported Ag 3d-M$_{45}$NN APECS measurements \cite{arena2000method} and shown how coincidence spectroscopy can be used to determine the intrinsic shapes of overlapping spectral lines. 
In their experiment, they measured the coincidences with two cylindrical mirror electron analysers (CMA) of 1 eV instrumental resolution. 
One CMA was set to a fixed energy in the Auger region, while the second CMA scanned the core level spectrum and, vice versa,  one CMA was set to a fixed core level energy region, while the other CMA scanned the Auger spectrum.

In contrast to their setup, like many other coincidence setups nowadays, the CoESCA station can record a coincidence map and a corresponding map of accidental coincidences in one run, without changing the analyser settings or any other experimental parameter.
The best guess of the true coincidences dataset is derived in post-processing by a simple subtraction, see \myeqref{eq:CalcTrueMap}.
 For electron-electron coincidences from surfaces, the successful subtraction of the accidental coincidences contribution from the coincidence dataset has previously been demonstrated in \cite{FrankMapSubtraction2011}.

 For the measurements presented in this paper, we chose a photon energy of 700 eV and operated the beamline at a resolution of 100 meV.
The \artof 10k spectrometer was set to record the Auger spectrum in the range of $335-359$ eV kinetic energy, while the \artof 2 EW spectrometer was set to record the photoelectron spectra in the range of $318-330$ eV kinetic energy.
The total experimental resolution was around 0.6 eV, determined from fitting the PES lines.
The position of the Fermi edge was used to calibrate the energy axis for both spectrometers in post-processing. 
The silver single crystal sample, Ag(111) (99.999\%, MaTeK GmbH), was cleaned by repeated $Ar^+$ sputtering and annealing cycles until no signs of contamination of the Ag surface by other materials, especially carbon, nitrogen and oxygen were visible in the photoelectron spectra.

The photon flux was adjusted via the exit slit of the beamline to achieve singles count rates of \num{0.8e3} $s^{-1}$ for the Auger region and \num{1.0e3} $s^{-1}$ for the core level region.
The total acquisition time was 24h, which yielded 106124 coincidence counts from single pulses and 101504 accidental coincidence counts from cross-comparing two consecutive pulses, resulting in 55372 true coincident counts. 
This corresponds to a measured coincidence rate of $\nu_c=1.2 \; s^{-1}$, composed of  $\nu_t=0.64\;s^{-1}$ true and  $\nu_a=0.59 \; s^{-1}$ accidental coincidences and and accidental-to-true ratio of $\nu_a / \nu_t = 0.92$.
The photon pulse repetition rate was 1.25 MHz.

\figCoinAndAccMap
\figTrueMapWithSpecs

We were operating the \artof 2 EW analyser at its design limits and therefore, the transmission function along its energy window was not constant.
For the present analysis, it is sufficient to account for this in post-processing by correcting the spectral intensities with a linear transmission function such, that for the singles photoelectron spectra (PES) the ratio of M$_5$ peak area over M$_4$ peak area would match the literature value of 0.67 \cite{Mar81_Ag_M4toM5ratio}.
In a recent study, the potential to operate angular resolved time of flight spectrometers beyond the standard limits and extending the time of flight range which can be analysed has been shown \cite{extendedARTOFevalutation}.
The \artof 10k detector, used for measuring the Auger spectra, was operated well within its specifications, with a constant transmission function.

The data maps as acquired are shown in \fig{fig:rawmaps}.
\fig{fig:truemap} shows the true coincidence map. 

The coloured frames in the map mark the integration regions, used for extracting the Auger spectra obtained in coincidence with the 3d$_{3/2}$ (orange) and 3d$_{5/2}$ (blue) photoelectron peaks, respectively. 
These are plotted on top together with the full sum of the coincidence map. 
Also the singles spectrum is shown. 
It is clearly seen that the singles spectrum has a much larger background. 
This reduction of background is another advantage of recording the Auger spectrum in coincidence with the appropriate photoelectron region. 
The PES spectra are plotted on the right. 
The PES peaks have two equivalent denotations: 3d$_{3/2}$ or M$_4$ for the peak at $E_b = 373.8$ eV and 3d$_{5/2}$ or M$_5$ for the peak at $E_b = 367.8$ eV.

The spectrum obtained in coincidence with the 3d$_{5/2}$ photoelectron line is a pure M$_5$NN Auger spectrum. 
For the spectrum in coincidence with the 3d$_{3/2}$ photoelectron line the situation is, however, slightly different. 
The orange region in the coincidence map contains not only 3d$_{3/2}$ photoelectrons. 
There is also a contribution from 3d$_{5/2}$ photoelectrons, which have lost energy due to shake-up or inelastic loss processes. 
n the case of silver these additional excitations will be delocalised and will not significantly modify the 3d$_{5/2}$ decay spectrum. 
For this reason the spectrum obtained by integration of the orange region will contain a contribution which is identical to the M$_5$NN Auger spectrum. 
In order to estimate the intensity of this contribution we have made a numerical fit of the 3d photoelectron peaks. 
As a model for the inelastic tail of the spectra we have used a Shirley background. 
From the fit we can then determine the intensity of the tail of the 3d$_{5/2}$ spectrum in the integration region of the 3d$_{3/2}$ peak.  
We find that, for our experiment, this contribution amounts to about 6\% of the M$_4$ peak intensity and we use this as an upper limit for the M$_5$NN contribution to the M$_4$NN spectrum. 
Note, that the amount of this type of true coincident noise depends on the amount of inelastic scattered signal in the recorded true coincident spectrum and thus on the full width of the detection window in energy space. 
Therefore our value of 6\% is true only for this experiment.
Another source for such a contribution to the coincidence background could be due to a Coster-Kronig (CK) type decay. 
For silver, however, only M$_4$M$_5$O CK are energetically allowed, and the cross-sections for those are negligibly small.

The M$_4$NN spectrum is shown in the top panel of \fig{fig:specsAuger_cleanANDtwoholeaxis} before and after correction for the M$_5$NN contribution. 
It is clearly seen that the M$_5$NN contribution background is present on the low kinetic energy side of the main peak. The main M$_4$NN peak is therefore unaffected by the M$_5$ contribution, as a result of the good resolution of the CoESCA setup.
The correction is also very small and changes the appearance of the spectrum very little. 
Also the fraction of M$_5$ contribution depends on the resolution of the setup. 
If measured with lower resolution the relative amount of the M$_5$ contribution would be larger.

The bottom panel of \fig{fig:specsAuger_cleanANDtwoholeaxis} shows the measured and intrinsic M$_4$ and M$_5$ two-hole spectra.
These were obtained by summing the kinetic energies for the true coincidences, as described in the previous section. 
The data was then sorted in a histogram with equidistant bins of 500 meV.

\figSpecAugerAndTwoHoleAxis

\section{Summary \& Outlook}
The  APECS technique has been introduced and a detailed overview of previous experiments on this topic was given.
The experimental end station for APECS at the BESSY II synchrotron facility, the CoESCA station, has been described together with important basic considerations on the data acquisition and treatment procedure in an electron-electron coincidence experiment. 
Fundamental limits on acquisition times and Signal-to-Noise ratio have been derived.
We have presented the first APECS results using the CoESCA station on a clean  Ag(111) single crystal for measuring coincidences between the Ag 3d core levels (M$_4$ and M$_5$) and the corresponding Auger electrons from M$_{45}$NN Auger decays.
The acquired true coincidence spectrum for the M$_5$NN Auger decay is already intrinsic.
The acquired M$_4$NN Auger spectrum, however,  still contains contributions from the M$_5$NN decay. 
We have explained the origin of these contributions and presented a method to obtain an estimate of the clean, intrinsic M$_4$NN spectrum.
Furthermore, we extracted two-hole binding energy spectra for each Auger decay, M$_4$NN and M$_5$NN, respectively, from the true coincidences map. 
These spectra can be determined only from coincidence measurements and show the distribution of the total energy needed to create a coincident Auger electron -- Photoelectron pair, leaving two holes in the sample's electronic system.

Further recent results from the CoESCA station on Auger- and Photoelectron coincindences of molecular O$_2$ adsorbed on Ag(111) 
\cite{Joh20_AgO_coins}
and quantifying Ni L$_{2,3}$ core-hole relaxation pathways 
\cite{Bor20_Ni_coins}
have been submitted for publication.

\section{Acknowledgements}
This work has been supported by the European Research Council (FP7/2007-2013)/ERC grant agreement no. [321319], the Swedish Research Council and Carl Tryggers Foundation for Scientific research (CTS).

\bibliographystyle{elsarticle-num} 
\bibliography{coesca.bib}

\end{document}